\documentclass[letter,12pt]{article}
\pdfoutput=1
\usepackage{jcappub}
\usepackage{graphicx,amssymb,amsmath,color,bm}
\usepackage{hyperref}
\usepackage{subcaption}
\usepackage{float}

\newcommand{\beq}{\begin{equation}}
\newcommand{\eeq}{\end{equation}}
\def\bea{\begin{eqnarray}}
\def\eea{\end{eqnarray}}
\def\bmat{\begin{matrix}}
\def\emat{\end{matrix}}

\newcommand{\bei}{\begin{itemize}}
\newcommand{\eei}{\end{itemize}}

\newcommand{\Fig}[1]{Fig.~\ref{#1}}
\newcommand{\Eq}[1]{Eq.~(\ref{#1})}
\newcommand{\Sec}[1]{Sec.~\ref{#1}}

\def\={\,=\,}
\def\+{\,+\,}
\def\-{\,-\,}
\def\Msun{M_\odot}

\begin{document}

\title{
Solar Diffraction of LIGO-Band Gravitational Waves
}

\author[a,b]{Sunghoon Jung, }
\author[a]{Sungjung Kim}

\affiliation[a]{Center for Theoretical Physics, Department of Physics and Astronomy, Seoul National University, Seoul 08826, Korea}
\affiliation[b]{Astronomy Research Center, Seoul National University, Seoul 08826, Korea}

\emailAdd{sunghoonj@snu.ac.kr}
\emailAdd{sjkimphya@gmail.com}

\abstract{
We show that chirping gravitational waves in the LIGO frequency band $f=1 - 5000$ Hz can be gravitationally diffracted by the Sun, due to the coincidence of its Fresnel length $r_F \propto \sqrt{1\, {\rm AU}/f}$ and the solar radius $r_\odot$. This solar diffraction is detectable through its frequency-dependent amplification of the wave, albeit with low event rates. We also advocate that solar diffraction allows probing the inner solar profile with the chirping evolution of frequencies. Along the course, we develop diffractive lensing in terms of simple convergence and shear of a lens and emphasize the relevance of high-frequency regimes including merger and ringdown phases for detection. This work not only presents an interesting opportunity with ongoing and future LIGO-band missions but also develops the diffractive lensing of long-wavelength waves in the universe. A similar phenomenon can also help discover non-relativistic wave dark matter, as studied in a sequel. 
}

\maketitle

\section{Introduction and summary}

Gravitational waves (GWs) from binary mergers have distinguishing features from other waves in the universe. First of all, they have well-predicted waveforms chirping in frequency and time. Such characteristic waveforms allow them to be detected~\cite{TheLIGOScientific:2016agk}, but also contain extractable information about dark matter and dark energy~\cite{Schutz:1986gp,LIGOScientific:2017adf}. In addition, their wavelengths are astrophysical or cosmological in length so that wave properties in their propagation, scattering, and lensing lead to unique phenomena and opportunities.

GW's long wavelength is right to produce observable interference fringes from the lensing by primordial black holes of the solar mass range~\cite{Nakamura:1997sw,Jung:2017flg,Lai:2018rto,Christian:2018vsi,Diego:2019lcd} or by cosmic strings of the strong tension~\cite{Jung:2018kde}. Even more wave-like diffraction is found to be useful in probing galactic (sub)halos of the diffuse NFW mass profile~\cite{Dai:2018enj,Choi:2021bkx} and cosmological matter power spectrum~\cite{Takahashi:2005ug,Oguri:2020ldf}. In more general, exquisite sensitivities to the chirp mass, eccentricity, and the last stage of mergers also open up new probes of wave dark matter~\cite{Choi:2018axi}, compact dark objects~\cite{Giudice:2016zpa,Cardoso:2019rvt}, dark matter captures~\cite{Ellis:2017jgp}, sub-solar mass black holes~\cite{Barsanti:2021ydd,Shandera:2018xkn}, and non-standard gravitation theories.

In this work, based upon Ref.~\cite{Choi:2021bkx} and pioneering works by Takahashi et. al.~\cite{Takahashi:2003ix,Takahashi:2005ug}, we advocate a new opportunity with LIGO-band GWs -- gravitational diffraction by the Sun and related physics. Along the study, we also develop intuitions and calculations of gravitational diffraction, in terms of convergence and shear, which make various underlying physics and estimations easier and clearer. They are essential to GW phenomenology. But they can also be applied to various other long-wavelength waves in the universe. As a remark, a similar phenomenon can be used to probe non-relativistic wave dark matter, which will be studied in a sequel~\cite{jung:ta}. 

Here, we summarize the main points of this work, which will be detailed in \Sec{sec:diff}. 
\bei
\item First of all, LIGO-band GWs are diffracted by the Sun. The condition for diffraction (beginning of \Sec{sec:diff}) makes it clear the relation between the LIGO frequency band and the Sun. 
\item Second, solar diffraction is detectable through its frequency-dependent amplification. The frequency dependence is a generic property of diffraction, and the chirping GW waveform allows us to detect it. The frequency dependence makes the high-frequency regime crucial for detection even though SNR gain may be small there (end of \Sec{sec:res1}). Further, diffraction can be readily understood and estimated by our formalism in terms of convergence and shear (middle of \Sec{sec:diff} and demonstrated in \Sec{sec:res1}). 
\item Lastly, solar diffraction allows probing the inner profile of the Sun with the chirping evolution. Our formalism in terms of convergence and shear also gives insights on what this opportunity means (later part of \Sec{sec:diff}). 
\eei
These are all aided by our calculation method based on analytic continuation (\Sec{sec:full}), which is not only intuitive but also computationally fast and accurate.

With these main contents, we start by introducing solar lens and the lensing language in \Sec{sec:sunlens}, and finish by calculating detection prospects in later sections \ref{sec:lnp} and \ref{sec:results}. We also make brief comments on the prospects with outer/inner solar system missions.

\section{Preliminaries: Solar lensing} \label{sec:sunlens}

In this section, we introduce the Sun as a gravitational lens, using the quantities projected onto the two-dimensional (2D) lens plane perpendicular to the line of sight. Those 2D quantities are suitable both for the usual (geometrical-optics) lensing as well as our main topic -- the diffractive lensing by the Sun. We review the former in this section and diffraction in the next section.

In our study, the Sun is a gravitational lens of the chirping GWs produced from cosmological binary mergers, and the observer is on the Earth. As will be discussed, the Sun cannot be treated as a point lens, since the Einstein radius $r_E \propto \sqrt{d_L}$ in this work is not much larger than the solar radius $r_\odot$; $d_L \sim 1\, {\rm AU} \ll d_{S,LS}$ denote the angular-diameter distances to the lens, to the source, and between the lens and the source. Thus, its inner mass profile shall be taken into account, using the Standard Solar Model (SSM)~\cite{Vinyoles:2016djt}.

Gravitational lensing is described in terms of potentials and locations projected onto the 2D lens plane; for more details, see e.g. Refs.~\cite{Schneider:1992,Dodelson:2020}. 
The 2D mass profile is ($x \equiv r/r_\odot$ is a lens-plane distance from the lens center located at the origin)
\beq
\Sigma(x) \= r_\odot \int_{-\sqrt{1-x^2}}^{\sqrt{1-x^2}} dz \, \rho( z^2 +x^2).
\eeq
Then we define dimensionless 2D quantities ($d_{\rm eff} \equiv d_L d_{LS}/d_S \simeq d_L$ in this work)
\bea
\kappa(x) &\=& \frac{\Sigma(x)}{\Sigma_{\rm cr}} \= 4\pi d_{\rm eff} \Sigma(x) \,\simeq \, 4\pi d_L \Sigma(x)  \\
\overline{\kappa}(x) &\=& \frac{4\pi d_{\rm eff}}{\pi x^2} \int_0^x dx^\prime \, 2\pi x^\prime \Sigma(x^\prime) \,\simeq\, \frac{8 \pi d_L}{x^2} \int_0^x dx^\prime x^\prime \Sigma(x^\prime) \\
\gamma(x) &\=& \overline{\kappa}(x) - \kappa(x),
\label{eq:gamma} \eea
where the critical density $\Sigma_{\rm cr} = 1/4\pi d_{\rm eff}$.

The convergence, $\kappa(x)$, obviously measures the projected mass density at the point $x$, related to the 2D potential $\psi(x)$ by the 2D Poisson equation $\nabla^2 \psi(x) = 2\kappa(x)$. But, by Gauss' theorem or Birkhoff's theorem, a more relevant quantity is the average density $\overline{\kappa}(x)$ enclosed within the radial distance $x$.
The cylindrical enclosed mass is
\bea
M(x) &\=& 2\pi r_\odot^2 \int_0^x dy \,y \Sigma(y)  \= \pi r_\odot^2 x^2 \overline{\Sigma}(x)
\= \overline{\kappa}(x) \frac{r_\odot^2 x^2}{4 d_L},
\eea
or, $M(\theta) = \overline{\kappa}(\theta) \theta^2 d_L/4$ in terms of the angle $\theta = r_\odot x / d_L$. Thus, the deflection angle at the impact parameter $\theta$ in the usual geometrical optics limit is
\bea
\alpha(\theta) &\=& \frac{ 4 M(\theta) }{ d_L \theta } \= \overline{\kappa}(\theta)\, \theta.
\eea
The lens equation between the source and image positions, $x_s$ and $x_i$, is
\beq
x_s \= x_i - \overline{\kappa}(x_i) \, x_i.
\eeq
When $x_s=0$, i.e. the source is aligned with the lens, the Einstein ring is formed, with the Einstein radius $x_E$ satisfying
\beq
\overline{\kappa}(x_E) \= 1.
\eeq
In other words, when $\overline{\kappa}(x) \lesssim 1$ for all $x$, the lens system does not produce multiple images. This can be used to define the Einstein radius more generally in terms of the enclosed mass within it 
\beq
\theta_E(x) \= \sqrt{ 4 M(\theta_E)/d_L } \= \sqrt{\overline{\kappa}(\theta_E)} \, \theta_E.
\eeq
We have reviewed how usual geometrical-optics lensing can be described in terms of $\overline{\kappa}(x)$.
We will see in the next section that diffraction is also described by $\overline{\kappa}(x)$.

The shear, $\gamma(x)$, is responsible for the asymmetric distortion of an image. Combined with $\overline{\kappa}(x)$, it allows to map the mass profile of the lens, just as the lens galaxy profile is mapped with the background galaxy shape distortion. In the context of diffraction, alternative interpretation aided by the expression in \Eq{eq:gamma} is useful -- it measures the \emph{variation} of the density profile at $x$. We will see in the next section that the shear measures the frequency-dependent \emph{change} of the diffraction amplification.

\begin{figure}
\includegraphics[width=\linewidth]{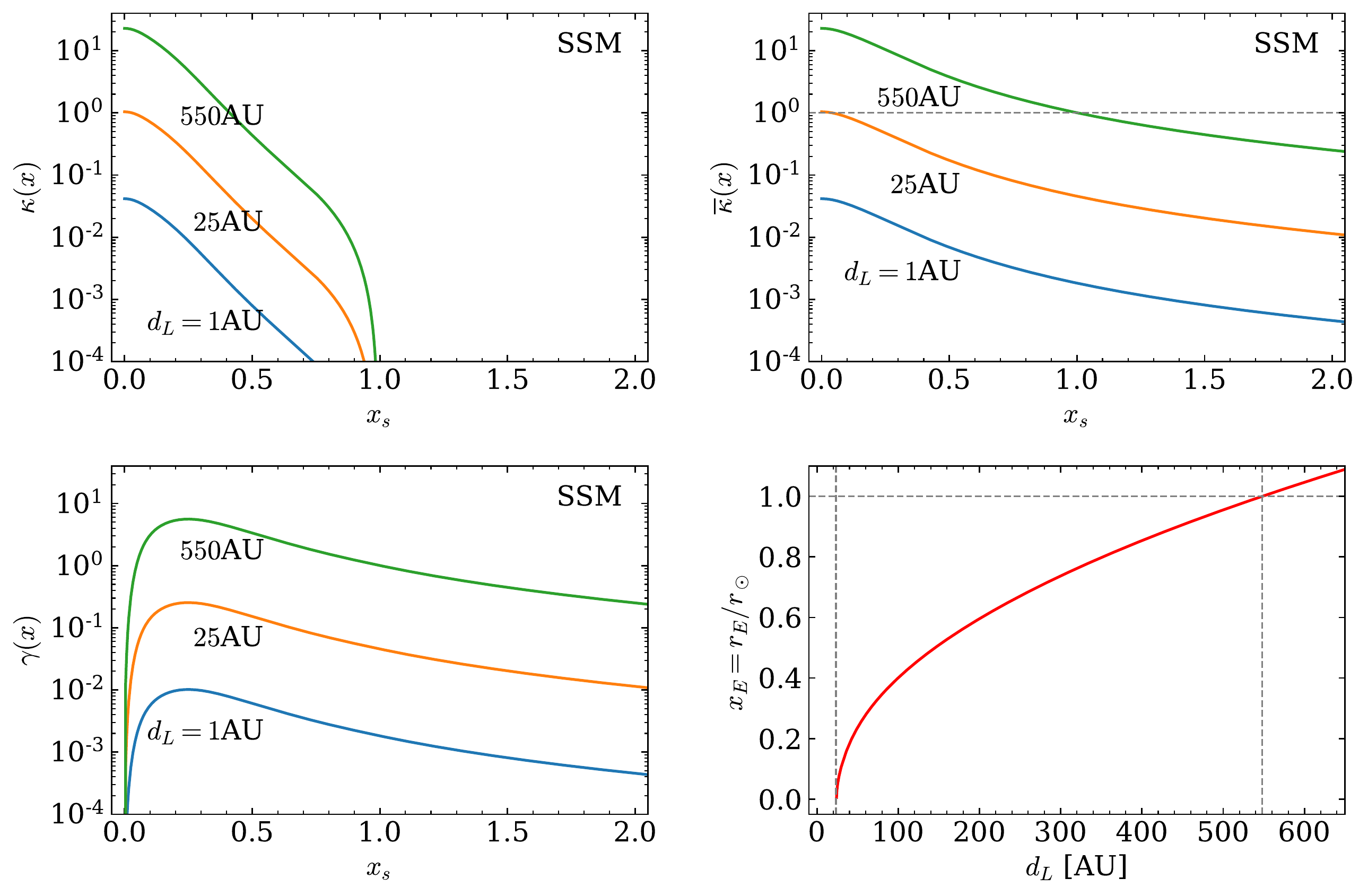}
\caption{ \label{fig:2D} 
2D projected lensing quantities of the Sun observed at $d_L = 1, 25, 550$ AU. The last panel shows the Einstein radius with $d_L$. The SSM solar mass profile~\cite{Vinyoles:2016djt}. 
}
\end{figure}

The 2D quantities for the Sun are shown in \Fig{fig:2D}. Obviously, they are the largest inside the Sun ($x<1$) and decreasing outward. 
Note that the 2D quantities are not the properties of the lens alone, but they are also proportional to $d_{\rm eff} \simeq d_L$. Thus lensing effects grow with $d_L$. It is because as $d_L$ is larger, small deflection at the lens plane can cause larger changes at the observer location. But probably the most interesting $d_L$ dependence is from the diffraction physics, as will be discussed in \Sec{sec:outer}.

Another notable is that $\overline{\kappa}(x) < 1$ for all $x$ if $d_L = 1$ AU, i.e. for the observer located on the Earth. Thus, the Sun will not produce multiple images when observed on the Earth. Only when $d_L \gtrsim 25$ AU, can multiple images be formed for some small $x_s$. The formation of multiple images can also be thought of as the focusing of two images at the location of the observer. Thus, the minimum focal length is $\sim 25$ AU~\cite{Patla:2007ju}, but this is possible only when the wave can pass through the Sun (small $x$). For the waves that cannot (such as light or neutrinos with some energy), the minimum focal length is $\sim 550$ AU, where all images form outside the Sun; it is also where the Einstein radius $r_E = r_\odot$.

When the Einstein radius is smaller than $r_\odot$ or is absent, the Sun cannot be treated as a point lens. The Einstein radius represents the boundary of strong lensing with multiple images; or the region with $\overline{\kappa} \gtrsim 1$. Thus, for $d_L =1$ AU as well as for observers located at the majority of outer solar systems, the Sun's mass profile has to be properly taken into account. The 2D projected quantities are useful for this, and the non-trivial radial mass profile (combined with the right wavelength of GW) is the one that produces interesting signals of solar diffraction.

As a side remark, these conclusions on the focal length and $r_E$ of the Sun change if the wave becomes non-relativistic with a small velocity. It is due to the Sommerfeld enhancement of gravitational scattering. For related interesting physics with axion-like wave dark matter, see the sequel~\cite{jung:ta}.

\section{Solar diffraction}   \label{sec:diff}

Waves are gravitationally diffracted (rather than deflected) when the Fresnel length $r_F$ matches with the characteristic length on the lens plane -- $r_s \sim r_\odot$ in this work. 
The Fresnel length is defined as~\cite{Takahashi:2003ix,Macquart:2004sh}
\beq
r_F \= \sqrt{ \frac{d_{\rm eff}}{2 \pi f}} 
\,\simeq\, 0.38 \, r_\odot \, \sqrt{ \left( \frac{d_L}{1\, {\rm AU}} \right) \left( \frac{100 \, {\rm Hz}}{f} \right) },
\label{eq:rF} \eeq
analogously to the single slit diffraction~\cite{Choi:2021bkx,Thorne:2017}. The slit of size $a$ is blurry imaged when the passing rays have small phase differences, $2\pi(\sqrt{a^2+d^2}-d)/\lambda \sim \pi a^2/(\lambda d) = (a/r_F)^2$.
Then the condition for diffraction is~\cite{Takahashi:2005ug,Oguri:2020ldf,Choi:2021bkx} 
\beq
r_F \gtrsim r_s,
\label{eq:diffcond} \eeq
meaning that the wavelength is too large to resolve the source location $r_s$ against the lens; for the single-slit case, $r_s$ corresponds to the slit size $a$. Similarly, $r_F$ can also be thought of as the effective source size~\cite{Oguri:2020ldf}.

\begin{figure}
\centering
\includegraphics[width=\linewidth]{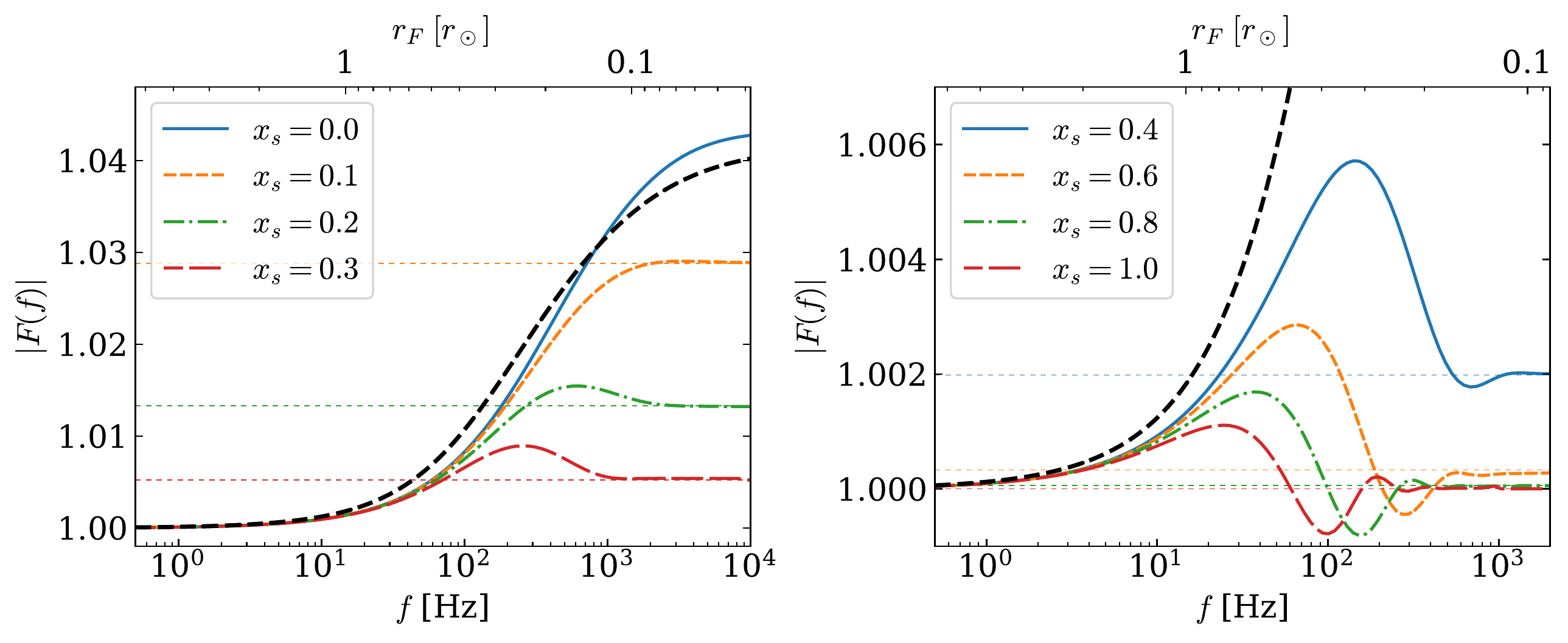}
\caption{ \label{fig:amp1} 
Solar diffraction amplification $|F(f)|$ for several source locations $x_s$. In addition to full results with analytic continuation, the back dashed line shows the approximate diffraction results in terms of $\overline{\kappa}(x)$ in \Eq{eq:amp-wave}; it does not depend on $x_s$ since it is not well resolved. They agree well at small frequencies where diffraction occurs, up to $f_{\rm trans}$ (\Eq{eq:diffgeo}) at which lensing transitions to the geometrical-optics (asymptotic values in \Eq{eq:amp-geo} are marked as horizontal dotted lines for each $x_s$). The transition accompanies oscillations for large $x_s$, which exhibit interference among diffracted waves. The frequency-dependent growth is most rapid in the LIGO frequency band $f=1-5000$ Hz, satisfying $r_F \sim r_s \sim r_\odot$ (\Eq{eq:diffcond}).}
\end{figure}

The diffraction amplification is approximately given by~\cite{Choi:2021bkx}
\beq
F(f) \,\simeq\, 1 + \overline{\kappa} \left(r \= r_F e^{i \frac{\pi}{4}} \right),
\label{eq:amp-wave} \eeq
where the lensing amplified wave amplitude is $\widetilde{h}_L(f) = F(f) \widetilde{h}(f)$. Diffraction is indeed measured by $\overline{\kappa}$, as alluded. Most importantly, it is frequency-dependent as explicit from the $r_F$ dependence, and the effect can be readily estimated by $\overline{\kappa}(x)$. But note that $\overline{\kappa}$ has to be analytically continued in this approximate result; the analytic continuation as well as full calculation will be discussed in \Sec{sec:full}. As shown in \Fig{fig:amp1}, \Eq{eq:amp-wave} describes diffraction well at low frequencies.

\begin{figure}
    \centering
    \includegraphics[width=\textwidth]{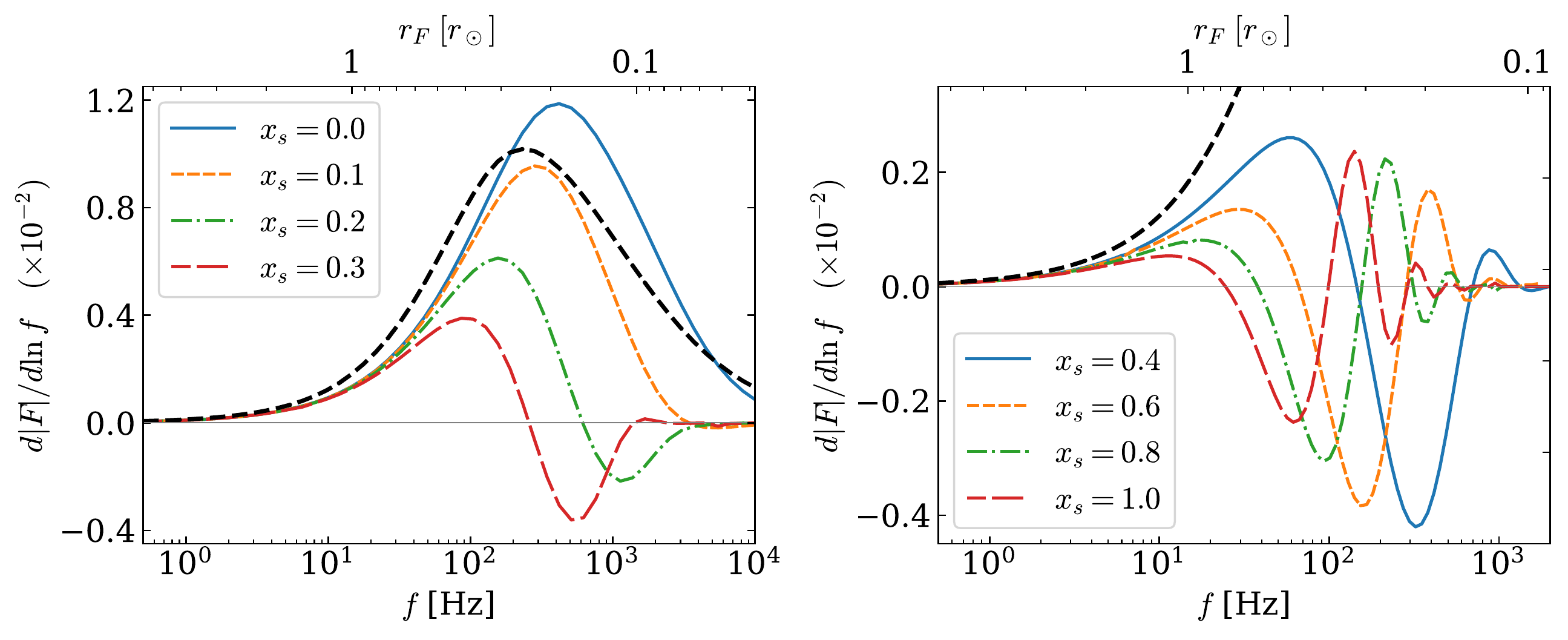}
    \caption{Same as \Fig{fig:amp1} but for $d|F|/d\ln f$. Approximate diffraction results in terms of $\gamma(x)$ in \Eq{eq:damp-wave} are overlaid as black dashed.
    }
    \label{fig:gam1}
\end{figure}

The frequency dependence in the diffraction regime is measured by the shear $\gamma(x)$ as~\cite{Choi:2021bkx}
\beq
    \frac{d|F(f)|}{d \ln f} \,\simeq\, \gamma\left( r\=r_F e^{i\frac{\pi}{4}} \right).
\label{eq:damp-wave}\eeq
This relation is consistent with \Eq{eq:amp-wave} and (\ref{eq:gamma}). Since lensing amplification measures the average density within $r_F$, its change with the frequency must measure the density variation at $r_F$, and this is nothing but the $\gamma(r=r_F)$ in \Eq{eq:gamma}; see more at the end of this subsection. As compared in \Fig{fig:gam1}, this approximate result agrees with full calculation at low frequencies.

At high frequencies (of chirping GWs) with $r_F \lesssim r_s$, diffraction approximations do not hold, and lensing moves into the geometrical optics regime, roughly at the frequency satisfying $r_F \sim r_s$
\beq
f_{\rm trans} \,\sim\, \frac{d_{\rm eff} }{ 2\pi r_s^2 } \,\simeq\, \frac{ d_L }{ 2\pi r_s^2 } \,\simeq\, 1500 \,{\rm Hz} \left(\frac{d_L}{1 \,{\rm AU}} \right) \left( \frac{0.1}{x_s} \right)^2.
\label{eq:diffgeo} \eeq
This rough estimate agrees with \Fig{fig:amp1} up to a factor $\sim 2$. The transition is not sharp, often accompanying oscillations that exhibit interference among diffracted waves; thus, oscillations are more prominent for larger $x_s$ as path length variations become larger. Also for the same reason, $f_{\rm trans}$ is smaller for larger $x_s$; the larger wavelength (hence a coarser probe) is enough to resolve the larger source separation.

Geometrical optics is where the wavelength is small enough to resolve the source away from the lens center so that the image becomes clear and frequency independent; $|F(f)|$ indeed becomes frequency independent in \Fig{fig:amp1}.
This frequency independent amplification is calculated by the magnification of the geometrical-optics image
\beq
F \,\simeq\, \sqrt{ \mu_i }  \= \sqrt{ \frac{1}{ (1-\kappa(x_i))^2 - \gamma(x_i)^2 } },
\label{eq:amp-geo} \eeq
where the location of the image is $x_i \simeq x_s$ for weak gravity, or more precisely satisfying the lens equation $x_s = x_i - \overline{\kappa}(x_i) x_i$. This assumes a single image, as $\overline{\kappa}<1$\footnote{For $d_L\gtrsim 25$ AU, $r_F \gtrsim r_s$ in the LIGO band. Thus, multi-image is not relevant to our study.}.
This geometrical-optics amplification roughly meets the diffraction result in \Eq{eq:amp-wave} at a frequency around \Eq{eq:diffgeo}, as shown in \Fig{fig:amp1}.

\medskip
Now we detail three main points of this work.
The first is that the LIGO-band frequencies $f=1-5000$ Hz of chirping GWs happen to be right to experience solar diffraction. It stems from the coincidence of scales $r_F \sim r_\odot$ (\Eq{eq:rF}) in this frequency band.

Secondly, solar diffraction is frequency-dependent, which makes it detectable with chirping GWs.
As chirping waveforms are characteristic in the frequency domain, additional effects from solar diffraction can be well discriminated. The same physics have been exploited to search for primordial black holes~\cite{Nakamura:1997sw,Jung:2017flg,Lai:2018rto,Christian:2018vsi,Diego:2019lcd}, cosmic strings~\cite{Jung:2018kde}, diffuse subhalo~\cite{Dai:2018enj,Choi:2021bkx}, and matter power spectrum~\cite{Takahashi:2005ug,Oguri:2020ldf}. 

In the rest of the work, we calculate the detectability of solar diffraction with LIGO-band missions. But one more exciting prospect remains to be discussed -- seeing or probing the inner solar profile using diffraction.

What do we mean by seeing inside the Sun? 
A notable property of approximate results \Eq{eq:amp-wave} and \Eq{eq:damp-wave} is that they depend on the 2D quantities evaluated at a particular location $r= r_F$, which is frequency dependent. Although a whole region on the lens plane is involved in the full calculation with path integral, the region around $r=r_F$ dominates the integral for diffraction. Strictly speaking, this is true only if the integration on the real line is rotated by $\pi/4$ to the complex plane~\cite{Choi:2021bkx}; the complex argument $e^{i\pi/4}$ makes it explicit. But the phase does not significantly change the qualitative description here. Thus, with the phase ignored, the diffraction of the frequency $f$ measures the density enclosed within $r_F \propto f^{-1/2}$ (\Eq{eq:amp-wave}), and the frequency-dependent change of diffraction must measure the density variation at $r_F$ (\Eq{eq:damp-wave}). As chirping sweeps a range of frequencies, the mass profile at successively smaller corresponding radii can be measured. This is what we mean by seeing inside the Sun. The same physics was exploited to probe diffuse NFW dark matter profile with GWs~\cite{Choi:2021bkx}.
Other ways to measure the lens mass profile include the well-known weak lensing by a galaxy and the eclipse by a lens star~\cite{Marchant:2019swq}.

\subsection{Full calculation with analytic continuation} \label{sec:full}

So far, we have discussed using approximate results. In our numerical results, we fully calculate the amplification factor $F(f)$. This is technically challenging, and here we describe our method. Along the way, an analytic continuation of 2D quantities will also be described. One can skip this subsection if not interested in technical details.

Lensing is calculated essentially by the path integral between the source and the observer. With the thin lens approximation, the integral can be taken over the 2D lens plane coordinate $\bm{x}$ as (also known as Kirchhoff integral)
\beq
    F(w) \= \frac{w}{2\pi i}\int d^2\bm{x} \, \exp[\, iw\hat{T}_d(\bm{x}, \bm{x}_s) \,],
    \label{eq:normalized_amp}
\eeq
where dimensionless quantities (frequency, time delay, and potential in order) are
\begin{align}
    &w \= 2\pi f\frac{r_\odot^2}{d_{\rm eff}}  \= \frac{r_\odot^2}{r_F^2},\\
    &\hat{T}_d \= \frac{1}{2}\left| \bm{x}-\bm{x}_s \right|^2 - \psi(\bm{x}), \\
    &\nabla^2_{\bm{x}} \psi \= 2\kappa(\bm{x}).
    \label{eq:psi_kappa}
\end{align}
For axisymmetric lenses, 
\beq
F(w) \= \frac{w}{i} \int_0^\infty dx\, x \exp\left[iw\frac{x^2-x_s^2}{2}\right] \, \exp[-iw\psi(x)] \, J_0(w x_s x) , \label{eq:amp_orig} 
\eeq
where $J_0$ is a Bessel function of the first kind of order zero.
This integral is hard to calculate even numerically. A fast Fourier transform (FFT) method was developed~\cite{ulmer1994femtolensing, nakamura1999wave}, but it takes huge computational time.

Instead, it is possible to analytically continue \Eq{eq:amp_orig} and rotate the integration real line by $\pi/4$ in the complex plane (hence, $x= y e^{i\pi/4}$)
\begin{align}
    F(w) \= e^{-iwx_s^2/2} w \int_0^\infty dy\, y \exp\left[ -w \frac{y^2}{2} \right] \, \exp\left[ -iw\psi(\sqrt{i}y ) \right] \, J_0(\sqrt{i}w x_s y) .
    \footnotemark
    \label{eq:amp_anal}
\end{align}
\footnotetext{Ordinary computational methods may produce overflows in the Bessel function when $\sqrt{2}r_s^2/r_F^2\gtrsim 10^3$, corresponding to the geometric optics regime.
Yet, the integral is safely evaluated in the diffraction regime, which we are mainly interested in.}
 One origin of the difficulties is that the integrand of \Eq{eq:amp_orig} oscillates in $x$, not converging to zero as $x\rightarrow \infty$. But the analytic continuation in \Eq{eq:amp_anal} makes the integral dominated by a local region around $y=1/\sqrt{w} = r_F/r_\odot$. Then, by using the Born approximation for weak gravity and ignoring $x_s$ for diffraction, we arrive at our intuitive results for diffraction in \Eq{eq:amp-wave} and (\ref{eq:damp-wave})~\cite{Choi:2021bkx}. 
 
\begin{figure}
        \centering
        \includegraphics[width=\textwidth]{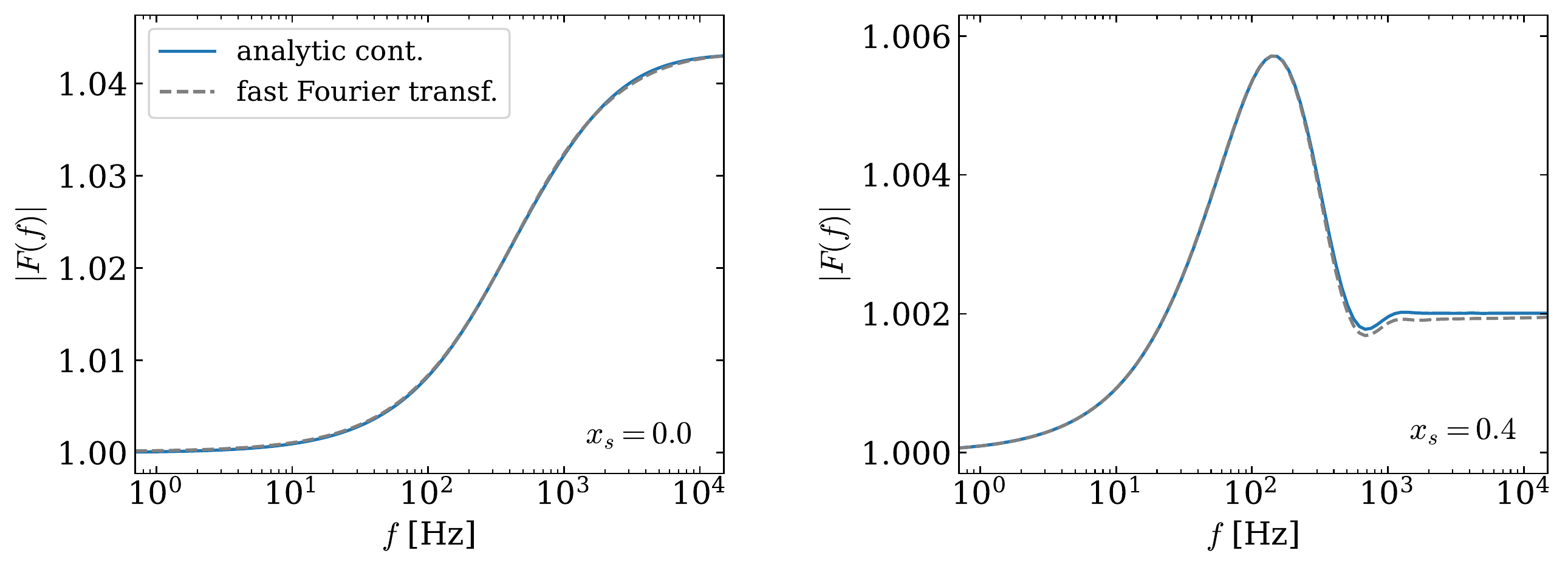}
    \caption{Comparison of full calculations of $F(f)$ using our analytic continuation (\Eq{eq:amp_anal}) versus fast Fourier transform (FFT). The difference between them is always $\mathcal{O}(0.01)\%$, agreeing well within unshown numerical errors; dominant errors come from FFT results in the low frequency and from the choice of analytic functional form in the high frequency.
    }
    \label{fig:amp_anal}
\end{figure}

To evaluate \Eq{eq:amp_anal} accurately, we need 
an analytic form of $\psi(x)$ in the complex plane. We use the following ansatz to fit the numerical data of SSM on the real line $z=x$ (the function should not have poles in $0\leq {\rm arg}(z) \leq \pi/4$)
\begin{align}
    \overline{\kappa}(z) = \frac{\overline{\kappa}(r=r_\odot)}{z^2+\dfrac{a}{b+z^c}},
\label{eq:kapan}\end{align}
and found the coefficients $a=2.9178\times10^{-3}$, $b=6.6307\times10^{-2}$, and $c=2.8755$. For large $z > 1$, $\overline{\kappa} \propto 1/z^2$ as it should be, but for small $z$, the $a$-term fits the solar profile inside.
Then, the solution of the Poisson equation \Eq{eq:psi_kappa} for the axisymmetric case, 
\begin{align}
    \psi(x) = \int_0^x dx'\, x'\overline{\kappa}(x'),
\end{align}
allows finding the analytic form of $\psi(x)$. 
These analytic fitting functions of $\overline{\kappa}(x)$ and $\psi(x)$ have at most 0.5\% differences with the SSM data on the real line. More importantly, as compared in Fig.~\ref{fig:amp_anal}, the analytic continuation yields only $\mathcal{O}(0.01)\%$ difference in $|F(f)|$ from the FFT results, while significantly reducing computation time. We use this analytic continuation method for our full calculation.

\section{Detection likelihood} \label{sec:lnp}

The main property of solar diffraction that allows its detection is its non-trivial frequency dependence.
It cannot be readily mimicked by general relativistic effects of binary mergers (such as spin, eccentricity, higher-order effects).
Exploiting this, we take a simplified approach to estimate the detection likelihood: maximizing the matching between the lensed waveform and the unlensed template by varying only the overall amplitude, phase, and time shift.
These variations do not induce extra non-trivial frequency dependences.
The detection likelihood in terms of these variables is known to approximate full estimates reasonably well~\cite{Choi:2021bkx,Dai:2018enj,Jung:2017flg,Christian:2018vsi}.

The template waveform at the leading order is
\begin{align}
    \widetilde{h}(f) \= A^0 A_p A(f) e^{i(2\pi f t_c^0 +\phi_c^0 + \Psi(f))}, \qquad \qquad {\rm (template)}
\end{align}
with $A(f)$ containing the chirping evolution in inspiral, merger, and ringdown phases
\begin{align}
    A(f) \= \left\{ \begin{matrix}
    A_{\rm insp}(f) && f < f_{\rm merg} && {\rm (inspiral)} \\
    A(f_{\rm merg}) \, \left( \frac{f}{f_{\rm merg}} \right)^{-2/3}  && f_{\rm merg} < f < f_{\rm ring} && {\rm (merger)} \\
    A(f_{\rm ring}) \, \frac{\sigma_f^2/4}{(f-f_{\rm ring})^2+\sigma_f^2/4}   && f_{\rm ring} < f < f_{\rm cut} && {\rm (ringdown)}  
\end{matrix} \right. ,
\end{align}
where
\begin{align}
    A_{\rm insp} (f) \= \sqrt{\frac{5}{96}} \frac{ {\cal M}^{5/6} f^{-7/6} \pi^{-2/3} }{D_s}. 
\end{align}
$D_s$ is the luminosity distance to the source.
Expressions for $f_{\rm merg, \,ring, \, cut}$ and $\sigma_f$ are collected in \cite{Ajith:2007kx}, but $f_{\rm merg}$ is approximately given by the innermost stable circular orbit 
\beq
f_{\rm merg} \,\sim\, \frac{1}{3\sqrt{3}\pi (M_1+M_2)} \,\simeq\, 207 \,{\rm Hz} \, \left( \frac{60 \, M_\odot}{M_1+M_2} \right)
\label{eq:fisco}\eeq
and it marks the end of the inspiral phase.
$A_p$ is a shorthand for binary and detector parameters (polarization, binary inclination, and detector antenna direction). Instead of specifying $A_p$, we normalize the amplitude to achieve the maximum SNR for the given distance, which is taken from the reported horizon distances~\cite{Abbott:2016xvh,Maggiore:2019uih,ETdesign}. For example, the maximum SNR at $D_s=400$ Mpc is 92 and 2150, with aLIGO and ET respectively.
$\Psi(f)$ is the chirping phase evolution.
It is canceled out in a likelihood measure between lensed and unlensed waveforms in \Eq{eq:lnp_base}.

The superscripted variables $A^0, t_c^0, \phi_c^0$ are fitting variables; lensed true waveforms are the ones without these parameters but multiplied by amplification $|F(f)|$. They represent the overall scaling, the shift of coalescence time, and the shift of the overall phase. As discussed, they do not induce extra non-trivial frequency dependencies.

The likelihood of lensing detection is measured by a $p$ value
\beq
    \ln p \= -\frac{1}{2} \,( h_L - h_{\rm BF} | h_L - h_{\rm BF} ),
    \label{eq:lnp_base}
\eeq
where the inner product is
\beq
(a|b) \= 4 {\rm Re} \int df \, \frac{ \widetilde{a}^*(f) \widetilde{b}(f) }{ S_n(f) }.
\eeq
The signal-to-noise ratio (SNR) is then $\rho^2 = (h|h)$.
$h_L$ is the lensed waveform
\beq
    \widetilde{h}_L(f) \= \widetilde{h}(f) F(f),
\eeq
and $h_{\rm BF}$ is the best-fitted waveform above minimizing $|\ln p|$.
The larger $|\ln p|$ means the worse fitting by unlensed waveforms or more confident detection of lensing.
We require $3\sigma$ significance for detection, corresponding to $\ln p \leq -5.914$.

\cite{Choi:2021bkx,Dai:2018enj} showed that the minimization with respect to $A^0$ and $\phi_c^0$ can be done analytically, yielding a useful form
\beq
    \ln p \= -\frac{1}{2} ( \rho_L^2 - \rho_{uL}^2 )
\label{eq:lnp2} \eeq
with SNR-squareds being
\beq
    \rho_0^2 \= (h_0 | h_0 ),
\eeq
\beq
    \rho_L^2 \= (h_L | h_L), 
\eeq
\beq
    \rho_{uL}^2 = \max_{t_c^0} \left| \frac{4}{\rho_0} \int df \, \frac{ |\widetilde{h}_0(f)|^2 }{S_n(f)} F^*(f) e^{2\pi i f t_c^0} \right|^2 .
\label{eq:rhoul} \eeq
The maximization with respect to $t_c^0$ barely changes $\rho_{uL}^2$ compared to the one with mere $t_c^0\= 0$.
So it was thought to be unimportant in \cite{Choi:2021bkx}. But the differences between $\rho_0^2$, $\rho_L^2$, and $\rho_{uL}^2$ are very small in the weak lensing.
Thus, we found a small but non-negligible change of $\ln p$ by the maximization with $t_c^0$.

Lastly, we note a useful property: $\ln p \propto \rho^2 \propto 1/D_s^2$ for the given observed mass $M$. We use $D_s = 400$ Mpc for benchmark discussions, but final results are obtained with full calculation.

\section{Result}  \label{sec:results}

\subsection{Detection likelihood of solar diffraction} \label{sec:res1}

\begin{figure}
    \centering
    \includegraphics[width=\textwidth]{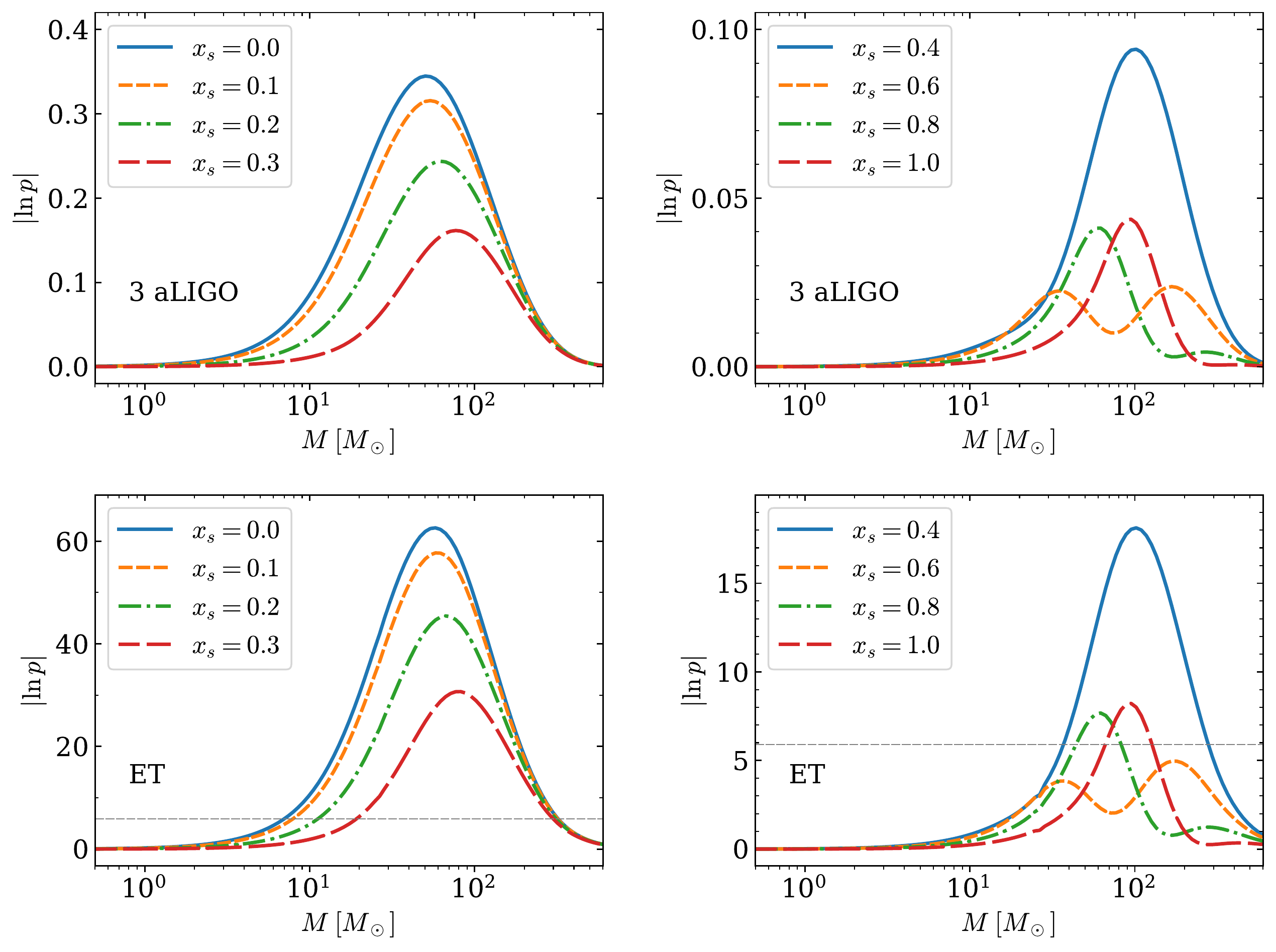}
    
    \caption{
        $|\ln p|$ likelihood of solar diffraction detection at aLIGO network (upper) and ET (lower). $D_s=400$ Mpc and $\ln p \propto 1/D_s^2$.
        Horizontal dashed lines denote the 3$\sigma$ detection threshold.
        $|\ln p|$ decreases smoothly with $x_s$ for small $x_s$, but fluctuates for large $x_s$.
    }
    \label{fig:lnp}
\end{figure}

Fig.~\ref{fig:lnp} shows $\ln p$ of solar diffraction with a network of three advanced LIGOs with design sensitivities (aLIGO)~\cite{TheLIGOScientific:2016agk} or with one Einstein Telescope (ET)~\cite{Punturo:2010zz,Hild:2010id}, for equal-mass binaries with $M_1=M_2=M$ (these masses denote the redshifted observed values) and fixed $D_s = 400$ Mpc. The noise curves are shown in \Fig{fig:spec} in Appendix~\ref{app:misc}.
Above all, $\ln p$ at aLIGO does not exceed $3\sigma$ significance at $D_s=400$ Mpc. 
On the other hand, $\ln p$ at ET can readily exceed the $3\sigma$ threshold over a range of masses $M=10 \sim 300 M_\odot$ and $x_s \lesssim 0.5$ for $D_s =400$ Mpc. With $\ln p \propto 1/D_s^2$, ET can perhaps detect up to $D_s \sim 4.2$ Gpc.
Let us understand these likelihoods in more detail using ET results; the detection event rates will be calculated in the next subsection.

The ET's sensitivity mass range is where corresponding chirping GWs merge in the LIGO frequency band; see $f_{\rm merg}$ in \Eq{eq:fisco} as a proxy, and also \Fig{fig:spec} for example waveforms. $M\sim 80 M_\odot$ is most sensitive in this range. GWs from heavier masses merge at lower frequencies so that a shorter frequency range can only be measured. Recall that a wide frequency range needs to be measured to detect the frequency-dependent effects of solar diffraction. On the other hand, GWs from lighter masses are simply too weak (small SNR) to be precisely measured. These two basic features are combined to yield $M \sim 80 M_\odot$ as the most sensitive mass range.

The detection likelihood is maximal for small $x_s \sim 0$ and decreases with $x_s$. Obviously, gravity is strongest inside. ET is expected to detect solar diffraction up to about $x_s \lesssim 0.5$ for $D_s =400$ Mpc. But for larger $x_s$, $\ln p$ does not monotonically decrease but often oscillates with $x_s$ as well as with $M$; see also \Fig{fig:lnpxs}. This is where interference becomes more rapid so that the oscillation of $|F(f)|$ (\Fig{fig:amp1}) becomes included in the measured frequency spectrum.
In any case, the maximum detectable $x_s$ is $\lesssim 1$ in the majority of parameter space, meaning that only GWs overlapped with the Sun can be strongly diffracted. The probability of detectable solar diffraction will thus be small of order $(x_s r_\odot / d_L)^2 / 4\pi \sim 10^{-6}$ for $D_s = 400$ Mpc.

\begin{figure}
    \centering
    \includegraphics[width=\textwidth]{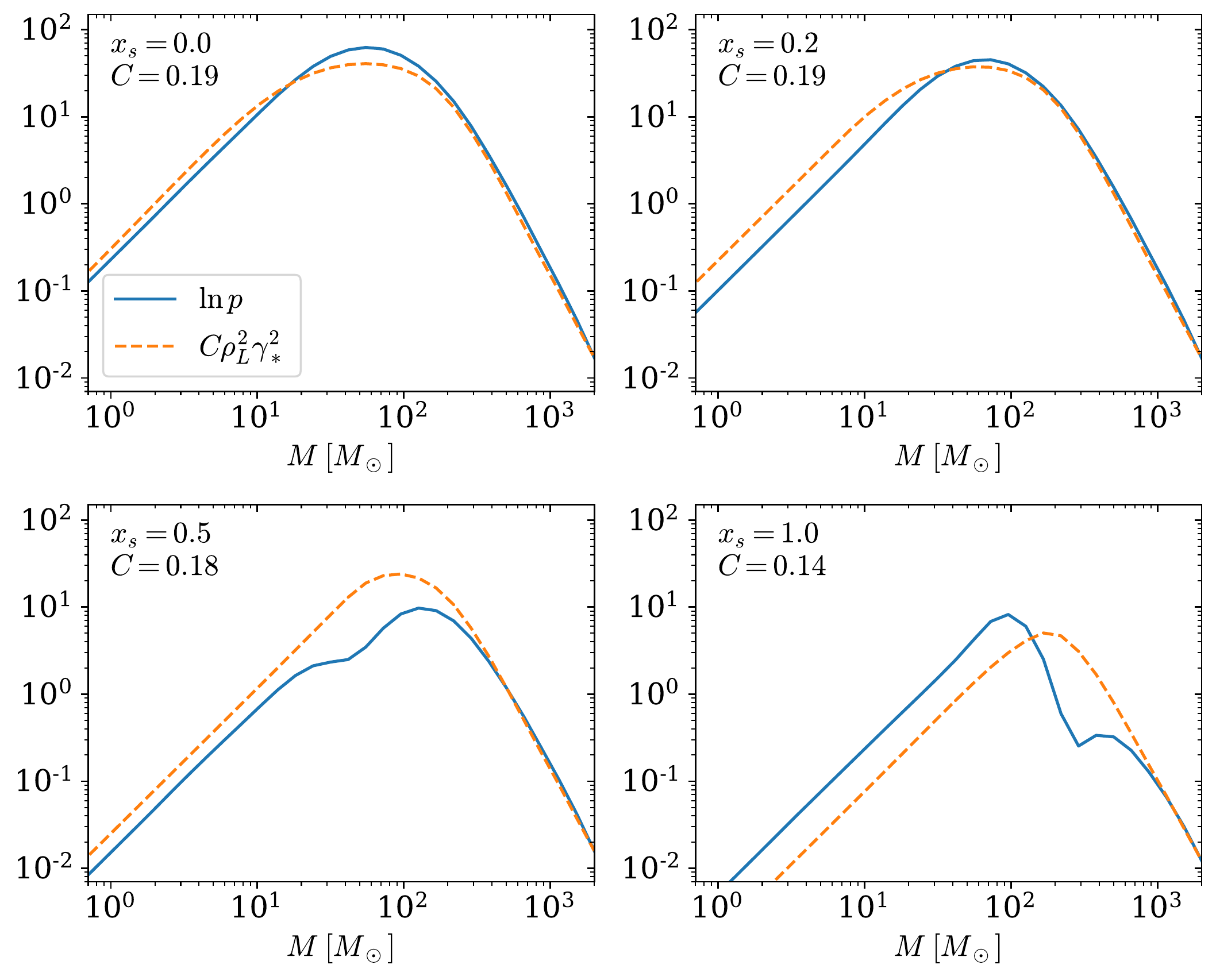}
    \caption{
    $\ln p$ estimates using diffraction approximation $\sim \rho_L^2 \gamma_*^2$ (dashed) is compared with full calculation (solid) at ET.
    $f_*$ is suitably chosen so that $|F(f_*)|-1$ is half of the maximum.
    The normalization factor $C = 0.14\sim 0.19$ is chosen to match the two results at $M=2000M_\odot$ for each $x_s$. See discussions in the main text.
    }
    \label{fig:gamma_star}
\end{figure}

\medskip
The $\ln p$ results can be further understood both quantitatively and intuitively.
It is  worthwhile to emphasize again that the main physics of detection is the frequency dependence of solar diffraction and this is measured by the shear $\gamma$ (\Eq{eq:damp-wave}).

Focus on the diffraction regime where $|F(f)|$ smoothly grows with $f$ (see \Fig{fig:amp1}). In this frequency range around $f_*$, the amplification can be approximated linearly as
\beq
    |F(f)| \,\simeq\, |F(f_*)| + \left. \frac{d|F(f)|}{d\ln f}\right|_* \Delta \ln f \,\sim\, |F(f_*)| + \mathcal{O}(1) \gamma_* ,
    \label{eq:amp_linear}
\eeq
where $\gamma_* \equiv \gamma(r_F(f_*))$ is the shear at the Fresnel length with $f=f_*$. This simplifies the comparison: the lensed waveform linearly grows in $f$ with the constant slope $\gamma_*$, while the unlensed waveform is flat with $F=1$ constant. $\ln p$ is essentially the integration of the difference squared (\Eq{eq:lnp_base}), so that~\cite{Choi:2021bkx} 
\begin{align}
    |\ln p| \,\sim\, \rho_L^2 \gamma_*^2 \mathcal{O}(1).
    \label{eq:lnp_estimation}
\end{align}
\Fig{fig:gamma_star} indeed shows that this estimation of $\ln p$ agrees well with full results (up to the ${\cal O}(1)$ factor, so the shape agrees well) for small $x_s$ and large $M$. These are the cases where only smoothly growing diffraction parts (with small $f$) are involved in the measured spectrum. When $x_s$ is large or $M$ is small, the high-frequency  geometrical-optics regime is involved in the measurements, and this linear approximation does not hold. The breakdown appears as extra features in the intermediate mass range in the figure. But the overall scaling with SNR still remains common, producing the same slopes for small $M$.

The bottom lines are that the \emph{observable} diffraction effect is its frequency dependency, and it is induced by the shear or the \emph{variation} of the mass profile, not just by the convergence or the density itself.
Furthermore, we can estimate the detection likelihood without dedicated analysis; the required SNR for detection is $\rho \gtrsim {\cal O}(1)/\gamma_*$.

\begin{figure}
    \centering
    \includegraphics[width=\textwidth]{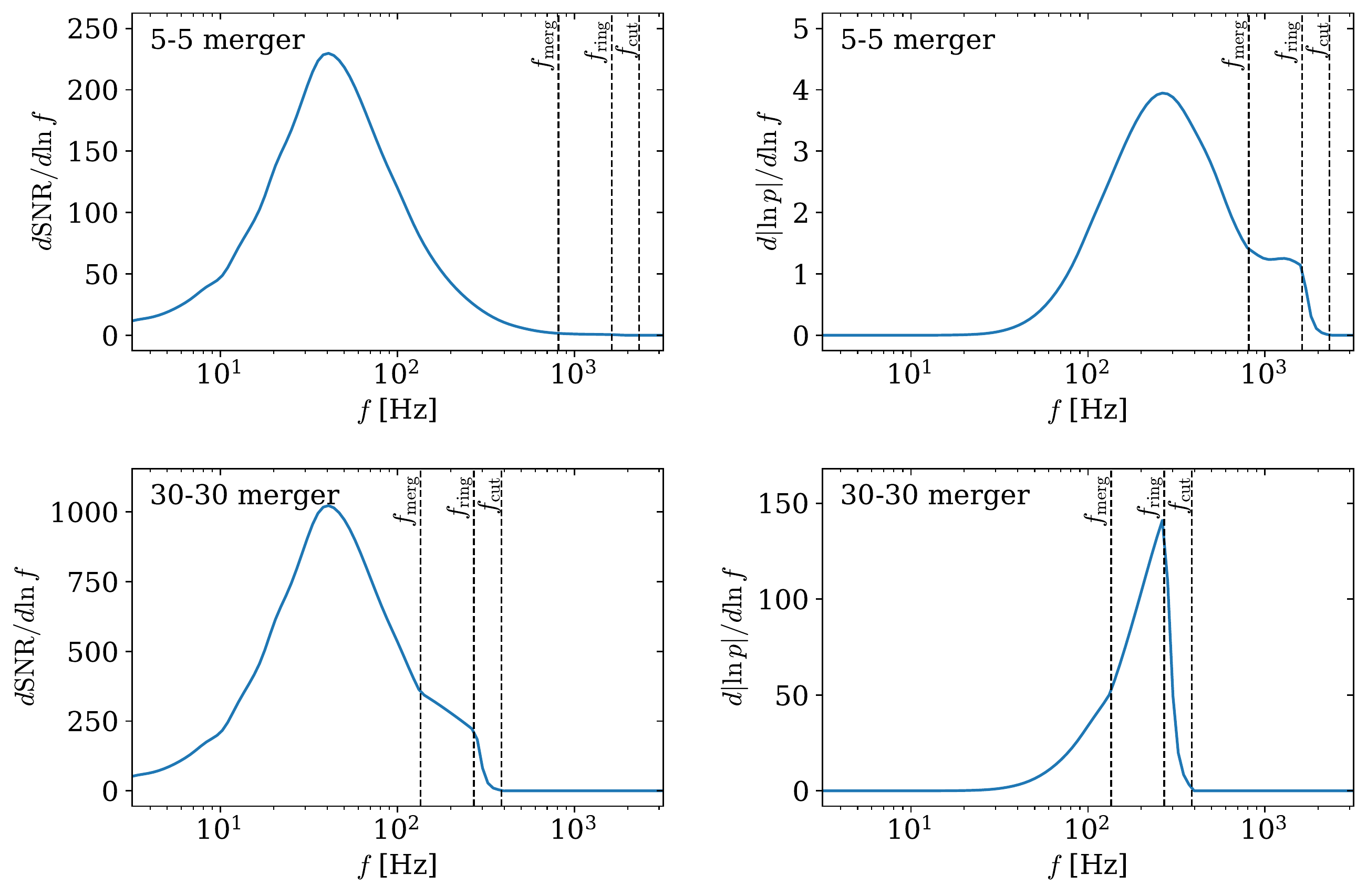}
    \caption{The accumulation rate of SNR (left) and $|\ln p|$ (right) in the log frequency interval as chirping proceeds. $x_s=0$ with ET. $M=5 M_\odot$ (upper) and $30 M_\odot$ (lower).}
    \label{fig:d_dlnf}
\end{figure}

\medskip
Another important lesson is that high-frequency regimes (including merger and ringdown phases) are crucial for diffraction detection, even though SNR gain might be small there. It can be learned by examining how $\ln p$ really captures the main physics of frequency dependencies. \Fig{fig:d_dlnf} shows that SNR and $\ln p$ are accumulated differently as chirping proceeds (shown  as the corresponding frequency in the $x$-axis); $\ln p$ is accumulated more at higher frequencies while SNR is more at lower frequencies where the noise is smaller.

The main difference is that SNR is accumulated locally while $\ln p$ only by comparison with other frequencies (hence globally). 
If one considers a measurement over a narrow frequency range, the SNR integrated over this range remains the same while $\ln p$ is much smaller, because no comparison can be made to detect frequency-dependent effects. Technically, the constant scaling $A^0$ can fit the diffraction enhancement in this narrow range alone, but it cannot fit a broader range measurement altogether.

In particular for large $M$ which merges at a smaller frequency, most $\ln p$ is accumulated at the highest frequency that it spends $\sim f_{\rm ring}$. For example, with only the inspiral phase (without merger and ringdown), $\ln p$ for $M=30 \Msun$ would have been smaller by a factor 4.9 and 4.5, respectively at aLIGO and ET. For $M=5 \Msun$, the impact is smaller.

If a broader frequency range is the only advantage, a small SNR gain there would not improve detection significantly. But it is the accumulation of frequency-dependent effects that makes the high-frequency regime crucial.
The relevance of high-frequency regimes for diffraction detection was also pointed out in \cite{Caliskan:2022hbu} also including spin effects.
The importance of high-frequency regimes also means that combining lower-frequency measurements from other missions would not help much. 

\subsection{Event rate}

\begin{figure}
    \centering
    \includegraphics[width=0.6\textwidth]{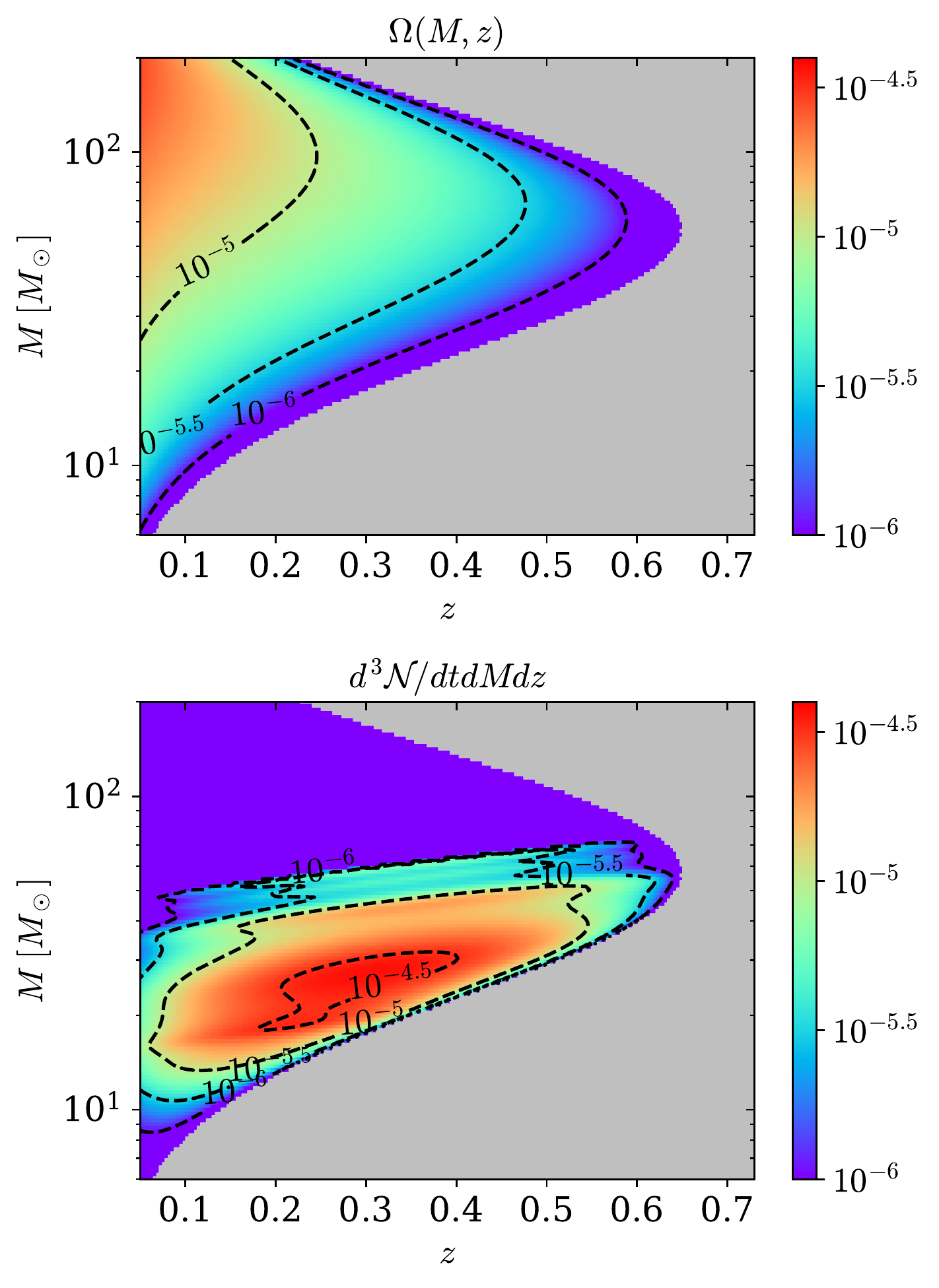}
    \caption{
        Detectable angular area $\Omega(z,M)$ (top) and the event rate density  $d^3\mathcal{N}/dt dM dz$ (bottom) with ET. The gray region has less than $3\sigma$ significance of detection and is the same in both panels. Any region smaller than $10^{-6}$ is colored purple.
    }
    \label{fig:omega_rate}
\end{figure}

The detectable GW diffraction event rate is
\begin{align}
    \frac{d\mathcal{N}}{dt} \= \int dM\,dz\,\frac{d\chi(z)}{dz}4\pi\chi^2(z)\frac{\Omega(M, z)}{4\pi}\frac{d^3\mathcal{N}}{dV_c dt dM}.
    \label{eq:detect_diff_num1}
\end{align}
$\mathcal{N}$ is the total number of detectable diffraction events (with more than $3\sigma$ significance), $t$ is the observation time, $\chi(z)$ is the comoving distance at redshift $z$, and $dV_c$ is a comoving volume element.
The distribution of binary mergers $d^3\mathcal{N}/dV_c dt dM$ is taken from~\cite{van2022redshift}; the total merger rate increases from $73 \,{\rm Gpc}^{-3} {\rm yr}^{-1}$ at $z=0$ to $155 \,{\rm Gpc}^{-3} {\rm yr}^{-1}$ at $z=0.7$.
$\Omega(M, z)$ is the angular area of the detectable source location $x_s$ for given $M$ and $z$, which is found from the maximum detectable $x_s$\footnote{Sometimes, small separated regions at large $x_s$ can be detectable due to oscillations in \Fig{fig:lnp} and \ref{fig:lnpxs}. However, those regions exist only when $D_s$ is small and have small contributions to the event rate, so we ignore them.}
In the event rate calculation, all information about diffraction is encoded in $\Omega(M, z)$.

\Fig{fig:omega_rate} shows $\Omega(M,z)$ and the event rate density $d^3\mathcal{N}/dt dM dz$ with ET. Above all, $\Omega \sim 10^{-6}$, as expected from the maximum detectable $x_s \lesssim 1$ in the previous subsection; thus it is roughly the angular area of the Sun on the sky. Secondly, the two observables exhibit different distributions.
As $z$ increases, the volume element quickly increases, making the merger density distribution tend toward large $z$. But too large $z$ reduces $\Omega(M,z)$. 
The turning point is at $z \simeq 0.3$, where the detectable event rate is maximum.
Also, the event rate density tends toward the small $M$ mainly because merger rates are higher at small masses.

The total detectable solar diffraction event rate with ET is $2.67 \times 10^{-4} \, {\rm yr}^{-1}$ and is very small with the aLIGO network. There exists a large uncertainty from merger rate calculations. By referring to the natal kick effects in Ref.~\cite{belczynski2016first}, we approximately take optimistic (pessimistic) merger rates to be 4 times larger (smaller) than those we take above from \cite{van2022redshift}.
The resulting optimistic and pessimistic event rates are $1.07\times 10^{-3} \, {\rm yr}^{-1}$ and $6.68\times 10^{-5} \, {\rm yr}^{-1}$, respectively.

In all, as usual, lensing event rates are limited by small lensing probabilities. Small GW merger rates further suppress in this case. If the detector sensitivity is $r$ times improved, the comoving horizon volume grows by $\sim r^3$, and the increase of $\Omega$ and the larger merger rate at large $z$ (at least up to $z \lesssim 7$~\cite{van2022redshift, belczynski2016first}) also enhance the event rate. Thus, if the next-generation GW sensitivity is improved by a factor $\sim 10$, several solar diffraction events may be detected every year.

\section{Prospects with solar-system missions} \label{sec:outer}

We have considered GW missions on the Earth, i.e. $d_L= 1$ AU. What about space missions farther out or inside the Solar system?

First of all, as discussed in \Sec{sec:sunlens}, lensing effects grow with $d_L$ as $\kappa, \overline{\kappa}, \gamma \propto d_L$ and $\ln p \propto \gamma_*^2 \propto d_L^2$ approximately. This may sound better in the outer solar system. But $\Omega$ and event rates do not necessarily grow with $d_L$. Although stronger lensing does increase the maximum detectable $x_s$, the corresponding angular area $\Omega\sim(x_s/d_L)^2$ does not necessarily increase with $d_L$. Thus, lensing probabilities and event rates are not necessarily improved in the outer solar system.

However, the change of $d_L$ does change the length scale. The Fresnel length $r_F \propto \sqrt{d_L}$ dictates the radial distance of the solar profile that can be probed with the diffraction of corresponding frequency $f$. Thus, in the inner/outer solar system, one may be able to probe the inner/outer part of the Sun better. 
This is an opportunity that becomes possible due to $d_{\rm eff}\sim d_L$, that we do not have with galactic dark matter subhalos, studied in \cite{Choi:2021bkx}.

\acknowledgments

We thank Han Gil Choi for useful discussions. We are supported by Grant Korea NRF-2019R1C1C1010050.

\appendix
\section{Supplementary figures} \label{app:misc}

\begin{figure}[h]
    \centering
    \includegraphics[width=.55\textwidth]{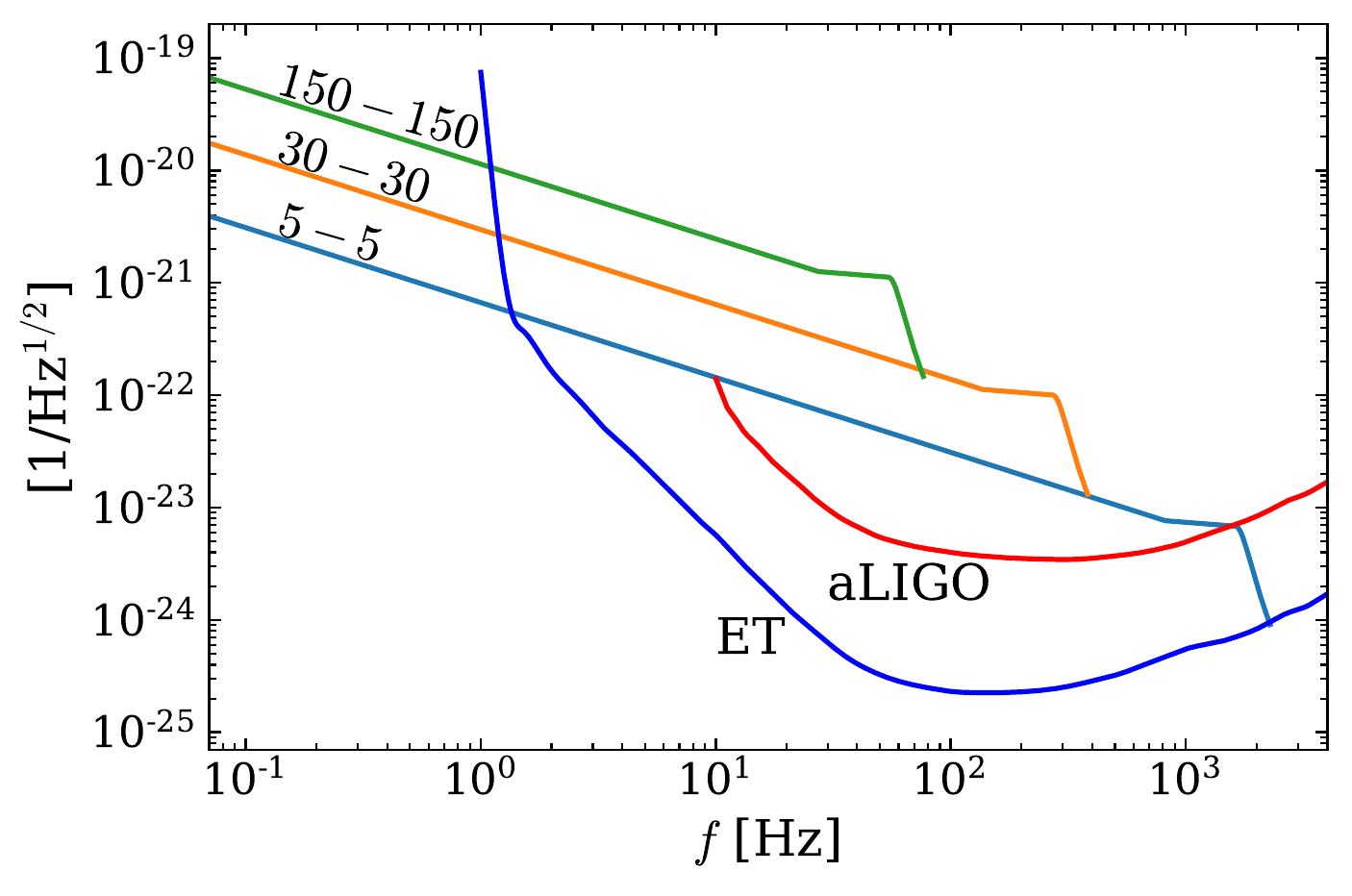}
    \caption{
    Strain sensitivities of aLIGO (design sensitivity)~\cite{TheLIGOScientific:2016agk} and ET~\cite{Punturo:2010zz,Hild:2010id} detectors, as well as GW strains of $M=150, 30, 5 \,M_\odot$ equal-mass binaries at $D_s =400$ Mpc. $A_p=1$.}
    \label{fig:spec}
\end{figure}

\begin{figure}[h]
    \centering
    \includegraphics[width=0.55\textwidth]{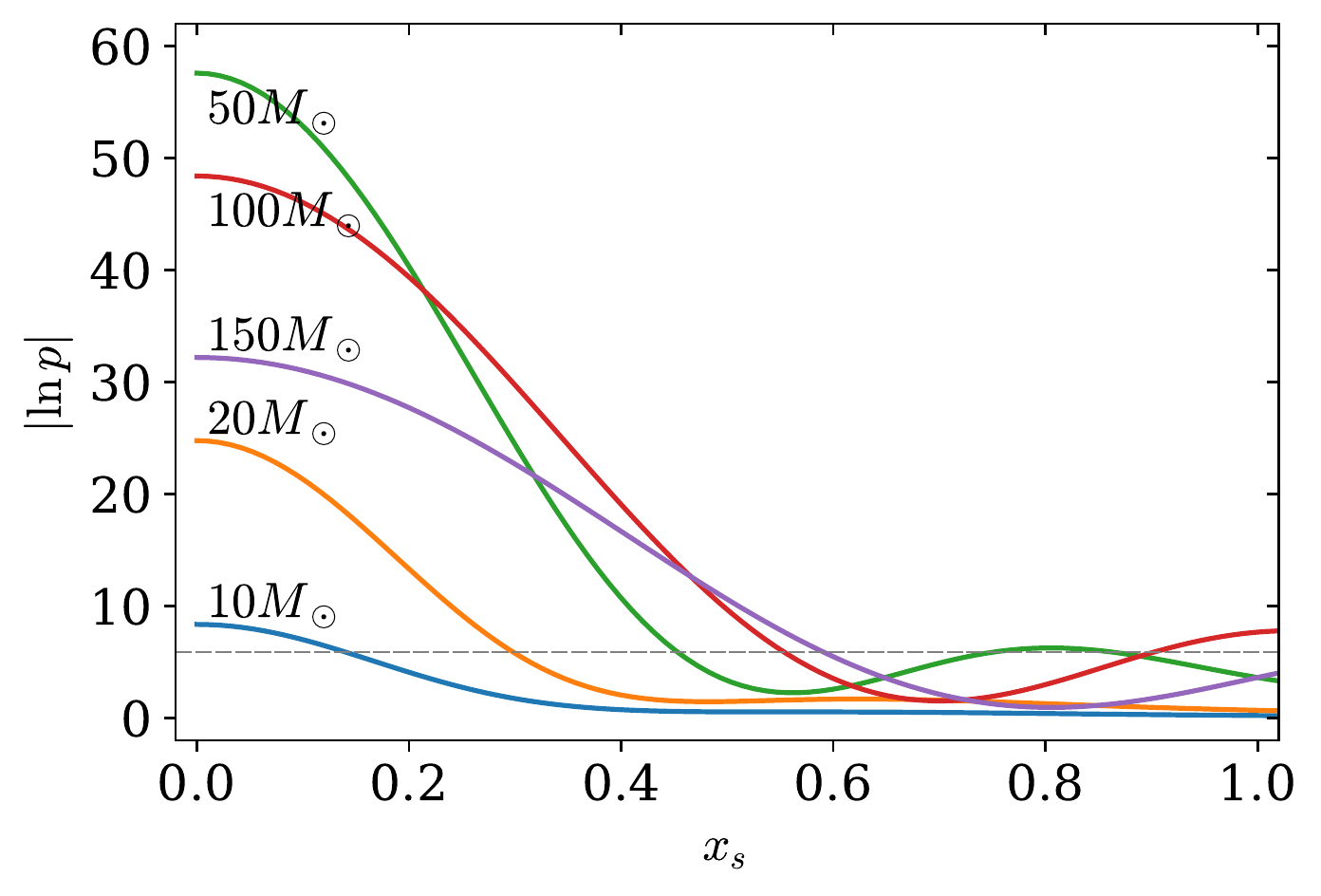}
    \caption{
        $\ln p$ at ET as a function of $x_s$ for several $M$ at $D_s=400$ Mpc.
        The horizontal dashed line denotes the $3\sigma$ threshold. It shows oscillations at large $x_s$ due to more rapid interference.
    }
    \label{fig:lnpxs}
\end{figure}


\begin{thebibliography}{99}


\bibitem{TheLIGOScientific:2016agk} 
  B.~P.~Abbott {\it et al.} [LIGO Scientific and Virgo Collaborations],
  ``GW150914: The Advanced LIGO Detectors in the Era of First Discoveries,''
  Phys.\ Rev.\ Lett.\  {\bf 116}, no. 13, 131103 (2016)
  doi:10.1103/PhysRevLett.116.131103
  [arXiv:1602.03838 [gr-qc]].


\bibitem{Schutz:1986gp}
B.~F.~Schutz,
``Determining the Hubble Constant from Gravitational Wave Observations,''
Nature \textbf{323}, 310-311 (1986)
doi:10.1038/323310a0

\bibitem{LIGOScientific:2017adf}
B.~P.~Abbott \textit{et al.} [LIGO Scientific, Virgo, 1M2H, Dark Energy Camera GW-E, DES, DLT40, Las Cumbres Observatory, VINROUGE and MASTER],
``A gravitational-wave standard siren measurement of the Hubble constant,''
Nature \textbf{551}, no.7678, 85-88 (2017)
doi:10.1038/nature24471
[arXiv:1710.05835 [astro-ph.CO]].

  
\bibitem{Nakamura:1997sw} 
  T.~T.~Nakamura,
  ``Gravitational lensing of gravitational waves from inspiraling binaries by a point mass lens,''
  Phys.\ Rev.\ Lett.\  {\bf 80}, 1138 (1998).
  doi:10.1103/PhysRevLett.80.1138

\bibitem{Jung:2017flg} 
  S.~Jung and C.~S.~Shin,
  ``Gravitational-Wave Fringes at LIGO: Detecting Compact Dark Matter by Gravitational Lensing,''
  Phys.\ Rev.\ Lett.\  {\bf 122}, no. 4, 041103 (2019)
  doi:10.1103/PhysRevLett.122.041103
  [arXiv:1712.01396 [astro-ph.CO]].

\bibitem{Lai:2018rto}
K.~H.~Lai, O.~A.~Hannuksela, A.~Herrera-Mart\'\i{}n, J.~M.~Diego, T.~Broadhurst and T.~G.~F.~Li,
``Discovering intermediate-mass black hole lenses through gravitational wave lensing,''
Phys. Rev. D \textbf{98} (2018) no.8, 083005
doi:10.1103/PhysRevD.98.083005
[arXiv:1801.07840 [gr-qc]].

\bibitem{Christian:2018vsi}
P.~Christian, S.~Vitale and A.~Loeb,
``Detecting Stellar Lensing of Gravitational Waves with Ground-Based Observatories,''
Phys. Rev. D \textbf{98} (2018) no.10, 103022
doi:10.1103/PhysRevD.98.103022
[arXiv:1802.02586 [astro-ph.HE]].

\bibitem{Diego:2019lcd}
J.~M.~Diego, O.~A.~Hannuksela, P.~L.~Kelly, T.~Broadhurst, K.~Kim, T.~G.~F.~Li, G.~F.~Smoot and G.~Pagano,
``Observational signatures of microlensing in gravitational waves at LIGO/Virgo frequencies,''
Astron. Astrophys. \textbf{627}, A130 (2019)
doi:10.1051/0004-6361/201935490
[arXiv:1903.04513 [astro-ph.CO]].

\bibitem{Jung:2018kde}
S.~Jung and T.~Kim,
``Probing Cosmic Strings with Gravitational-Wave Fringe,''
JCAP \textbf{07}, 068 (2020)
doi:10.1088/1475-7516/2020/07/068
[arXiv:1810.04172 [astro-ph.CO]].

\bibitem{Dai:2018enj}
L.~Dai, S.~S.~Li, B.~Zackay, S.~Mao and Y.~Lu,
``Detecting Lensing-Induced Diffraction in Astrophysical Gravitational Waves,''
Phys. Rev. D \textbf{98}, no.10, 104029 (2018)
doi:10.1103/PhysRevD.98.104029
[arXiv:1810.00003 [gr-qc]].

\bibitem{Choi:2021bkx}
H.~G.~Choi, C.~Park and S.~Jung,
``Small-scale shear: Peeling off diffuse subhalos with gravitational waves,''
Phys. Rev. D \textbf{104}, no.6, 063001 (2021)
doi:10.1103/PhysRevD.104.063001
[arXiv:2103.08618 [astro-ph.CO]].

\bibitem{Takahashi:2005ug}
R.~Takahashi,
``Amplitude and phase fluctuations for gravitational waves propagating through inhomogeneous mass distribution in the universe,''
Astrophys. J. \textbf{644}, 80--85 (2006)
doi:10.1086/503323
[arXiv:astro-ph/0511517 [astro-ph]].

\bibitem{Oguri:2020ldf}
M.~Oguri and R.~Takahashi,
``Probing Dark Low-mass Halos and Primordial Black Holes with Frequency-dependent Gravitational Lensing Dispersions of Gravitational Waves,''
Astrophys. J. \textbf{901} (2020) no.1, 58
doi:10.3847/1538-4357/abafab
[arXiv:2007.01936 [astro-ph.CO]].


\bibitem{Choi:2018axi}
H.~G.~Choi and S.~Jung,
``New probe of dark matter-induced fifth force with neutron star inspirals,''
Phys. Rev. D \textbf{99}, no.1, 015013 (2019)
doi:10.1103/PhysRevD.99.015013
[arXiv:1810.01421 [hep-ph]].

\bibitem{Giudice:2016zpa}
G.~F.~Giudice, M.~McCullough and A.~Urbano,
``Hunting for Dark Particles with Gravitational Waves,''
JCAP \textbf{10}, 001 (2016)
doi:10.1088/1475-7516/2016/10/001
[arXiv:1605.01209 [hep-ph]].

\bibitem{Cardoso:2019rvt}
V.~Cardoso and P.~Pani,
``Testing the nature of dark compact objects: a status report,''
Living Rev. Rel. \textbf{22}, no.1, 4 (2019)
doi:10.1007/s41114-019-0020-4
[arXiv:1904.05363 [gr-qc]].

\bibitem{Ellis:2017jgp}
J.~Ellis, A.~Hektor, G.~H\"utsi, K.~Kannike, L.~Marzola, M.~Raidal and V.~Vaskonen,
``Search for Dark Matter Effects on Gravitational Signals from Neutron Star Mergers,''
Phys. Lett. B \textbf{781}, 607-610 (2018)
doi:10.1016/j.physletb.2018.04.048
[arXiv:1710.05540 [astro-ph.CO]].

\bibitem{Shandera:2018xkn}
S.~Shandera, D.~Jeong and H.~S.~G.~Gebhardt,
``Gravitational Waves from Binary Mergers of Subsolar Mass Dark Black Holes,''
Phys. Rev. Lett. \textbf{120}, no.24, 241102 (2018)
doi:10.1103/PhysRevLett.120.241102
[arXiv:1802.08206 [astro-ph.CO]].

\bibitem{Barsanti:2021ydd}
S.~Barsanti, V.~De Luca, A.~Maselli and P.~Pani,
``Detecting Subsolar-Mass Primordial Black Holes in Extreme Mass-Ratio Inspirals with LISA and Einstein Telescope,''
Phys. Rev. Lett. \textbf{128}, no.11, 111104 (2022)
doi:10.1103/PhysRevLett.128.111104
[arXiv:2109.02170 [gr-qc]].



\bibitem{Takahashi:2003ix} 
  R.~Takahashi and T.~Nakamura,
  ``Wave effects in gravitational lensing of gravitational waves from chirping binaries,''
  Astrophys.\ J.\  {\bf 595}, 1039 (2003)
  doi:10.1086/377430
  [astro-ph/0305055].

  

\bibitem{Vinyoles:2016djt}
N.~Vinyoles, A.~M.~Serenelli, F.~L.~Villante, S.~Basu, J.~Bergstr\"om, M.~C.~Gonzalez-Garcia, M.~Maltoni, C.~Pe\~na-Garay and N.~Song,
``A new Generation of Standard Solar Models,''
Astrophys. J. \textbf{835}, no.2, 202 (2017)
doi:10.3847/1538-4357/835/2/202
[arXiv:1611.09867 [astro-ph.SR]].

\bibitem{Schneider:1992}
P.~Schneider, J.~Ehlers, E.~E.~Falco,
``Gravitational Lenses,'' (Springer, New York, 1992).

\bibitem{Dodelson:2020}
S.~Dodelson,
``Gravitational Lensing,'' (Cambridge, 2020).

\bibitem{Patla:2007ju}
B.~Patla and R.~J.~Nemiroff,
``Gravitational Lensing Characteristics of the Transparent Sun,''
Astrophys. J. \textbf{685}, 1297 (2008)
doi:10.1086/588805
[arXiv:0711.4811 [astro-ph]].



\bibitem{Macquart:2004sh}
J.~P.~Macquart,
``Scattering of gravitational radiation: Second order moments of the wave amplitude,''
Astron. Astrophys. \textbf{422}, 761--775 (2004)
doi:10.1051/0004-6361:20034512
[arXiv:astro-ph/0402661 [astro-ph]].

\bibitem{Thorne:2017}
 K.~S.~Thorne,  R.~D.~Blandford, 
 ``Modern Classical Physics: Optics,
Fluids, Plasmas, Elasticity, Relativity, and Statistical Physics'' (Princeton Univ. Press, 2017)


\bibitem{Marchant:2019swq}
P.~Marchant, K.~Breivik, C.~P.~L.~Berry, I.~Mandel and S.~L.~Larson,
``Eclipses of continuous gravitational waves as a probe of stellar structure,''
Phys. Rev. D \textbf{101}, no.2, 024039 (2020)
doi:10.1103/PhysRevD.101.024039
[arXiv:1912.04268 [astro-ph.SR]].



\bibitem{ulmer1994femtolensing}
A.~Ulmer and J.~Goodman,
``Femtolensing: Beyond the Semi-Classical Approximation,''
Astrophys. J. \textbf{442}, 67 (1995)
doi: 10.1086/175422

\bibitem{nakamura1999wave}
T.~Nakamura and S.~Deguchi,
``Wave optics in gravitational lensing,''
Prog. Theor. Phys. Suppl. \textbf{133}, 137-153 (1999)
doi: 10.1143/PTPS.133.137





\bibitem{Ajith:2007kx}
P.~Ajith \textit{et al.}
``A Template bank for gravitational waveforms from coalescing binary black holes. I. Non-spinning binaries,''
Phys. Rev. D \textbf{77}, 104017 (2008)
[erratum: Phys. Rev. D \textbf{79}, 129901 (2009)]
doi:10.1103/PhysRevD.77.104017
[arXiv:0710.2335 [gr-qc]].


\bibitem{Abbott:2016xvh}
B.~P.~Abbott, R.~Abbott, T.~D.~Abbott, C.~Adams, R.~X.~Adhikari, R.~A.~Anderson, S.~B.~Anderson, K.~Arai, M.~A.~Arain and S.~M.~Aston, \textit{et al.}
``Sensitivity of the Advanced LIGO detectors at the beginning of gravitational wave astronomy,''
Phys. Rev. D \textbf{93}, no.11, 112004 (2016)
doi:10.1103/PhysRevD.93.112004
[arXiv:1604.00439 [astro-ph.IM]].

\bibitem{Maggiore:2019uih}
M.~Maggiore, C.~Van Den Broeck, N.~Bartolo, E.~Belgacem, D.~Bertacca, M.~A.~Bizouard, M.~Branchesi, S.~Clesse, S.~Foffa and J.~Garc\'\i{}a-Bellido, \textit{et al.}
``Science Case for the Einstein Telescope,''
JCAP \textbf{03}, 050 (2020)
doi:10.1088/1475-7516/2020/03/050
[arXiv:1912.02622 [astro-ph.CO]].

\bibitem{ETdesign}
ET steering committee \textit{et al.}
``Einstein Telescope: Science Case, Design Study and Feasibility Report,''
available at https://apps.et-gw.eu/tds/ql/?c=15662



\bibitem{Punturo:2010zz}
M.~Punturo \textit{et al.}
``The Einstein Telescope: A third-generation gravitational wave observatory,''
Class. Quant. Grav. \textbf{27}, 194002 (2010)
doi:10.1088/0264-9381/27/19/194002

\bibitem{Hild:2010id} 
  S.~Hild {\it et al.},
  ``Sensitivity Studies for Third-Generation Gravitational Wave Observatories,''
  Class.\ Quant.\ Grav.\  {\bf 28}, 094013 (2011)
  doi:10.1088/0264-9381/28/9/094013
  [arXiv:1012.0908 [gr-qc]].

  
\bibitem{Caliskan:2022hbu}
M.~\c{C}al\i{}\c{s}kan, L.~Ji, R.~Cotesta, E.~Berti, M.~Kamionkowski and S.~Marsat,
``Observability of lensing of gravitational waves from massive black hole binaries with LISA,''
[arXiv:2206.02803 [astro-ph.CO]].


\bibitem{van2022redshift}
L.~A.~C.~van Son, S.~E.~de Mink, T.~Callister, S.~Justham, M.~Renzo, T.~Wagg, F.~S.~Broekgaarden, F.~Kummer, R.~Pakmor and I.~Mandel,
``The Redshift Evolution of the Binary Black Hole Merger Rate: A Weighty Matter,''
Astrophys. J. \textbf{931}, no.1, 17 (2022)
doi:10.3847/1538-4357/ac64a3
[arXiv:2110.01634 [astro-ph.HE]].

\bibitem{belczynski2016first}
K.~Belczynski, D.~E.~Holz, T.~Bulik and R.~O'Shaughnessy,
``The first gravitational-wave source from the isolated evolution of two 40-100 Msun stars,''
Nature \textbf{534}, 512 (2016)
doi:10.1038/nature18322
[arXiv:1602.04531 [astro-ph.HE]].



\bibitem{jung:ta} 
  H.~G.~Choi, S.~Kim and S.~Jung,
  To appear soon
  



  
\end{thebibliography}
\end{document}